\documentclass[12pt,preprint]{aastex}

\newcommand\msun{M_{\odot}}

\newcommand\msunyr{M_{\odot}\,\rm yr^{-1}}
\newcommand\be{\begin{equation}}
\newcommand\en{\end{equation}}

\newcommand\etal{{\rm et al}.\ }

\newcommand\mdot{\dot{M}}

\def\micron{$\mu$m}
\def\microns{$\mu$m }

\begin{document}

\shortauthors{Muzerolle et al.}
\shorttitle{Dynamical changes in a transitional disk}

\title{Evidence for Dynamical Changes in a Transitional Protoplanetary Disk
with Mid-infrared Variability}

\author{
James Muzerolle\altaffilmark{1,2,3},
Kevin Flaherty\altaffilmark{2},
Zoltan Balog\altaffilmark{2,4},
Elise Furlan\altaffilmark{5},
Paul S. Smith\altaffilmark{2},
Lori Allen\altaffilmark{6,7},
Nuria Calvet\altaffilmark{8},
Paola D'Alessio\altaffilmark{9},
S. Thomas Megeath\altaffilmark{10},
August Muench\altaffilmark{6},
George H. Rieke\altaffilmark{2},
William H. Sherry\altaffilmark{11}}

\altaffiltext{1}{Space Telescope Science Institute, 3700 San Martin Dr., Baltimore, MD 21218}
\altaffiltext{2}{Steward Observatory, 933 N. Cherry Ave., The University of Arizona, Tucson, AZ 85721}
\altaffiltext{3}{Visiting Astronomer at the Infrared Telescope Facility, which is operated by the University of Hawaii under Cooperative Agreement no. NCC 5-538 with the National Aeronautics and Space Administration, Science Mission Directorate, Planetary Astronomy Program.}
\altaffiltext{4}{Max-Planck-Institut f\"ur Astronomie, K\"onigstuhl 17, D-69117, Heidelberg, Germany}
\altaffiltext{5}{Jet Propulsion Laboratory, California Institute of Technology, Pasadena, CA 91109}
\altaffiltext{6}{Harvard-Smithsonian Center for Astrophysics, 60 Garden St., Cambridge, MA 02138}
\altaffiltext{7}{National Optical Astronomy Observatory, Tucson, AZ, 85719}
\altaffiltext{8}{Department of Astronomy, University of Michigan, Ann Arbor, MI 48109}
\altaffiltext{9}{Centro de Radioastronom\'ia y Astrof\'isica, UNAM, Morelia, Michoac\'an, M\'exico}
\altaffiltext{10}{Department of Physics and Astronomy, University of Toledo, Toledo, OH 43606}
\altaffiltext{11}{National Solar Observatory, Tucson, AZ 85719}

\begin{abstract}
We present multi-epoch Spitzer Space Telescope observations of
the transitional disk LRLL 31 in the 2-3 Myr-old star forming region IC 348.
Our measurements show remarkable
mid-infrared variability on timescales as short as one week.
The infrared continuum emission exhibits
systematic wavelength-dependent changes that suggest corresponding dynamical
changes in the inner disk structure and variable shadowing of outer disk
material.  We propose several possible sources for the structural changes,
including a variable accretion rate or a stellar or planetary
companion embedded in the disk.  Our results indicate that variability
studies in the infrared can provide important new constraints on protoplanetary
disk behavior.
\end{abstract}


\section{Introduction}

Circumstellar disks around young stars are the sites of planet formation.
Study of their structure and evolution is necessary
to understand planet formation mechanisms and constrain timescales.
The dust in a protoplanetary disk is heated by a combination of stellar
irradiation and viscous energy release; its emission provides important
clues to the overall structure of the disk.  Observations of
spectral energy distributions (SEDs) in the infrared trace
the thermal emission from this dust at distances of $\sim 0.1-10$ AU from
a young solar-type star, probing the main planet-forming regions of disks.

A substantial body of mid-infrared observations
(particularly from the Spitzer Space Telescope) has provided
statistically significant results on the overall frequency of protoplanetary
disks as a function of stellar age and mass (Lada et al. 2006; Hern\'andez
et al. 2007).  Furthermore, models of disk structure are increasingly
successful in reproducing time-averaged SEDs and spatially-resolved
scattered light images (e.g. Calvet et al. 2005; Pinte et al. 2008).
However, most of these studies assume a static, steady-state,
and symmetric structure for disks.  Observations
of optical and near-infrared variability indicate a more dynamic picture
of accretion activity (Carpenter et al. 2001; Eiroa et al. 2002; Bouvier
et al. 2007), and contradict
some of the underlying assumptions that are generally employed.
Observations of variability at mid-infrared wavelengths, which directly probe
thermal dust emission from protoplanetary disks, have been sparsely reported
(e.g. Liu et al. 1996; Barsony et al. 2005; Juh\'asz et al. 2007);
systematic studies are just beginning (Sitko et al. 2008;
\'Abrah\'am et al. 2009).  Here we report on the discovery of
surprising variations in the mid-infrared SED of the T Tauri star LRLL 31,
a member of the 2-3 Myr-old IC 348 star forming region and of particular
interest since it is surrounded by a transitional disk that likely has
a large inner hole or gap (Muzerolle et al. 2009).

\section{Observations}

\subsection{Spitzer}

We observed LRLL 31 multiple times over the span of about 6 months
with the Spitzer Space Telescope's IRS (Houck et al. 2004) and MIPS
(Rieke et al. 2004) instruments
to probe mid-infrared ($\lambda \sim 5$-40 \micron) variability
on timescales of days to months.
The observations were motivated by a comparison of two existing epochs
of Spitzer IRAC (Fazio et al. 2004) and MIPS imaging data separated by
about 5 months, which indicated a change in flux
at all wavelengths from 3.6 to 24 \micron.  We obtained two pairs of
IRS spectra (in staring mode) in fall 2007 and spring 2008,
with each pair separated by
one week.  We further obtained a total of 7 new MIPS observations,
each scanning the entire IC 348 cluster, with 5 taken on consecutive days
in fall 2007 and two more separated by one week in spring 2008.
All of these observations, including the previous Spitzer images and
ancillary ground-based data discussed below, are summarized in Table~\ref{obs}.

For the IRS observations, both Short-Low (SL) and Long-Low (LL) modules
were used in order to cover the full wavelength range, 5 to 40 $\mu$m.
Exposure times and number of cycles were 14x2 and 6x1 seconds for
Short-Low and Long-Low, respectively.  The spectra were extracted using
the Spectroscopic Modeling, Analysis, and Reduction Tool
(SMART;~Higdon et al. 2004) version 6.4, starting with the basic
calibrated data products from the Spitzer Science Center reduction
pipeline version S16.1.  Rogue pixels were first removed using the "IRS\_CLEAN"
program, constructing a pixel mask for each of the SL and LL modes using
off-order images as a reference.  SMART was then used with manual point source
extraction using a tapered column aperture and local sky subtraction.
The local sky background was calculated by fitting a first order polynomial
to the median of a group of 6 sky pixels on either side of the source at each
row in the spatial direction.
We opted not to use off-nod sky subtraction because the sky background
in the images was strong and non-uniform, causing residuals in the
subtracted images at longer wavelengths.
For comparison, we did extract spectra from off-nod subtracted images;
the results were identical within the formal uncertainties at all but
the longest wavelengths where the sky background is strongest,
with fluxes differing by $\lesssim 10\%$ at $\lambda \gtrsim 30$ \micron.
The final calibrated spectra for each nod and order
were then combined using the sigma-clipped averaging function in SMART.
We applied no offsets to individual orders during this step since they
were already aligned at a level within the signal-to-noise.
The formal measurement uncertainties propagated by SMART were typically
2\% over most of the spectrum, increasing to $\sim 5\%$ at
$\lambda \gtrsim 30$ \microns as a result of extra noise introduced
by the (subtracted) stronger sky background.

The MIPS observations were taken in scan mode using 12 scan legs with
30' length, half-array offsets, and medium scan rate, with coordinates
centered on the IC 348 cluster center (the same parameters used for the GTO
observations published in Lada et al. 2006).
We only analyzed the 24 \microns data for LRLL 31, since the exposure depth
and strong background emission at longer wavelengths precluded any useful
upper limit measurements at 70 and 160 \micron.
Starting with the raw data, individual images were calibrated and mosaicked
using the MIPS GTO team Data Analysis Tool (Gordon et al. 2005) version 3.06.
Photometry was measured using PSF fitting routines
in the IDLPHOT package.
We also re-reduced the existing GTO (PID 6 and 58) and c2d legacy
(PID 178) imaging data for IC 348.
The MIPS data were processed and analyzed as above.  For the IRAC data,
all frames were processed using the SSC IRAC Pipeline v14.0.0,
and mosaics were created from the basic calibrated
data (BCD) frames using a custom IDL program~(Gutermuth et al. 2008).
Aperture photometry on these images was carried out using PhotVis
version 1.10, which is an IDL-GUI based photometry visualization
tool~(Gutermuth et al. 2004).  The relative photometric uncertainties
are $< 1$\% for both IRAC and MIPS-24 (the {\it absolute} uncertainties
are larger).

\subsection{Ancillary Data}

We observed LRLL 31 with optical and near-infrared spectrographs in
order to characterize the stellar and accretion properties.
We obtained a near-infrared spectrum using SpeX~(Rayner et al. 2003)
on IRTF.  The spectrograph was in SXD mode,
covering a wavelength range of 0.8-2.5 \micron, with a 0.5" slit providing
spectral resolution of $\sim 1200$.  Data were taken with an ABBA slit-nodding
pattern, 120-second exposures at each nod, and 2 nod cycles for a total
on-source exposure time of 8 minutes.  The data were reduced using
the IDL-based SPEXTOOL reduction package~(Cushing et al. 2004).  

We also obtained a set of spectropolarimetric observations using the SPOL
imaging/ spectropolarimeter~(Schmidt et al. 1992) on the Steward Observatory
2.3 m telescope.  The first observation was made with the
instrument configured for spectropolarimetry.  A 600 l/mm diffraction
grating was used to yield a spectrum covering 4000-8000 \AA~with a
resolution of about 20 \AA~for the given slit width (3") employed.
The second SPOL observation used a plane mirror instead of a
grating to image a $\sim 1$x1 arcmin$^2$ field around LRLL 31.  In this case,
the bandpass was determined by the combination of Hoya Y48 and HA30
filters ($\sim 4800$-7000 \AA) and a 6"-diameter synthetic photometric aperture
was used for the polarization analysis.  Details of the
spectropolarimetric data acquisition and reduction can be found in,
e.g.,~Smith et al. 2003.

\section{Results}

\subsection{Stellar Properties}

We compared the near-infrared spectrum to a suite of SpeX spectral standards
from the IRTF spectral library (Rayner et al. 2009) to derive the spectral
type and visual extinction.  We derive a spectral type of G6,
with an uncertainty of about 2 subtypes, and a visual extinction
$A_V = 10$.  This spectral type is later, and the extinction
smaller, than previously reported (Luhman et al. 2003).  Using the pre-main
sequence tracks of Siess et al. (2000), we then estimate a stellar mass
of 1.8 $\msun$.  

The SpeX spectrum shows emission at the Paschen $\beta$ and
Brackett $\gamma$ hydrogen lines.  We measured line equivalent widths
for these features, after first subtracting a photospheric template in order
to account for the underlying photospheric absorption, of
$EW(Pa\beta) = -2.3$ \AA~and $EW(Br\gamma) = -1.8$ \AA.
These emission lines are known to be accurate tracers of accretion in
T Tauri stars, with well-defined relationships between line luminosity and
accretion luminosity ($L_{acc}$) (Muzerolle et al. 1998).  We calculated line
luminosities by scaling to the 2MASS J- and K-band magnitudes;
assuming $L_{acc}=0.8 \, GM_*\mdot/R_*$, we then derived
mass acccretion rates of
$\mdot=1.5 \times 10^{-8}, 1.7 \times 10^{-8} \, \msunyr$ from Pa $\beta$
and Br $\gamma$, respectively.
These estimates are roughly a factor of 5 lower than previously published
values from Dahm (2008) based on data taken about 11 months later.
Given the typical measurement uncertainties of 30\% or less,
LRLL 31 may thus have a time-variable $\mdot$ on timescales of one year or less.

From the SPOL spectropolarimetry and imaging, we measured optical
polarization percentages and position angles of
$P = 7.66 \pm 0.25$\%, $\theta = 144.0 \pm 0.9 \deg$ and
$P = 8.41 \pm 0.20$\%, $\theta = 146.4 \pm 0.7 \deg$, respectively.
The difference between the two measurements
may be a result of the different effective bandpasses rather than
intrinsic source variability.  The SPOL spectrum also provided information
on the stellar and accretion properties.  By fitting the optical integrated
light spectrum to a G6 stellar model, we derived a visual extinction of
$A_V \sim 8.3$.
The amount of polarization and the difference in extinction
derived from the near-infrared spectrum indicate a significant
amount of scattered light and suggest a highly inclined orientation for
the system and/or partial obscuration of the central star by material in
the circumstellar disk.

\subsection{Infrared variability}

\begin{figure}
\plotone{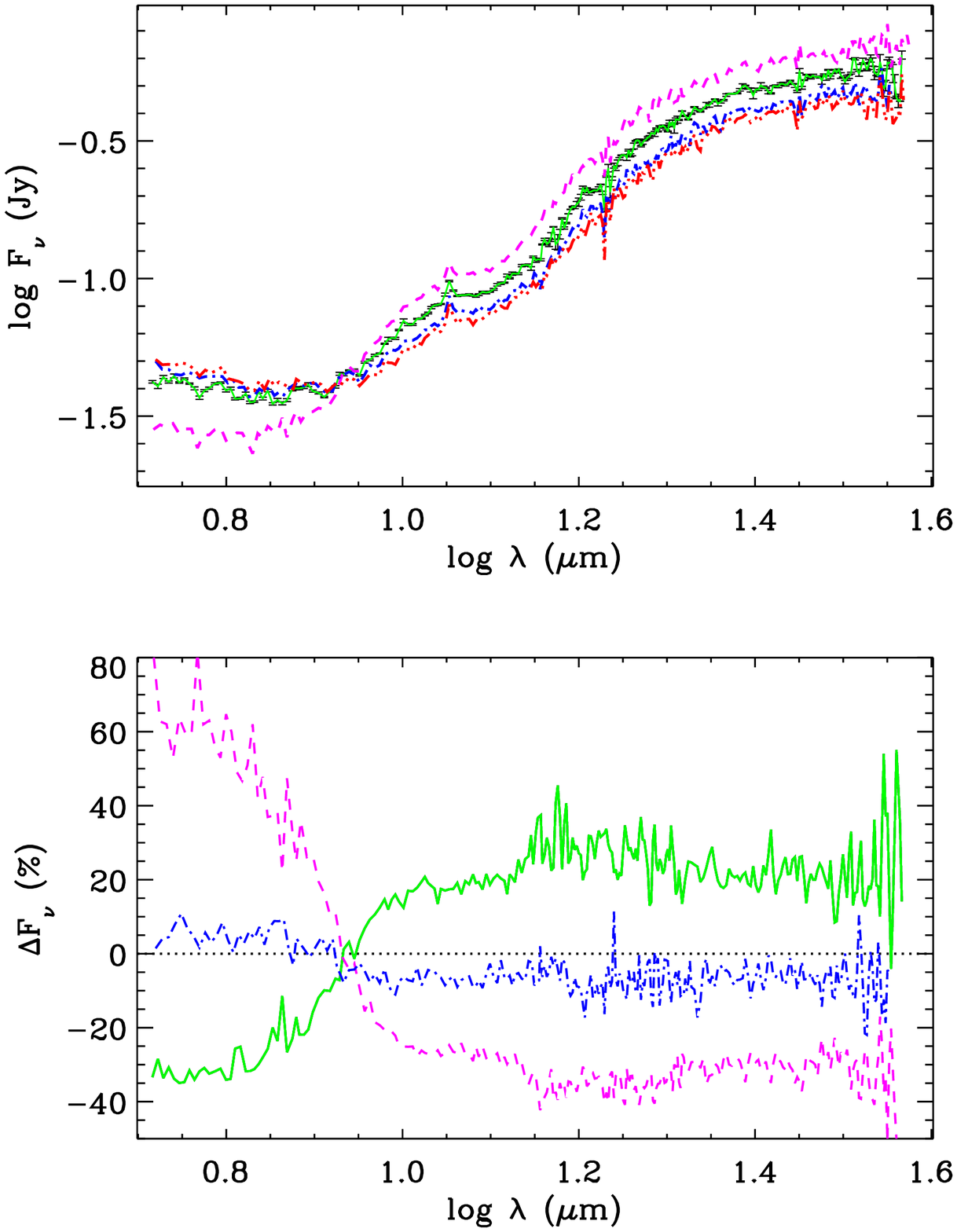}
\caption{Top: IRS spectra of LRLL 31, lines in the following order: solid green
(2007 Oct 9), dashed magenta (2007 Oct 16), dash-dot blue (2008 Feb 24), and
dash-triple-dot red (2008 Mar 2).  Error bars are indicated for the first epoch
spectrum.  Bottom: difference spectra between the first and second epochs
(solid green), second and third epochs (dashed magenta), and third and fourth
epochs (dash-dot blue), as a percentage change in flux.
\label{irs}}
\end{figure}

Figure~\ref{irs} shows the four epochs of IRS spectra taken for LRLL 31.
The top panel shows each spectrum in flux density units, while the bottom
panel shows the difference spectra derived by subtracting epochs 1 and 2,
2 and 3, and 3 and 4, respectively.  Each epoch exhibited flux changes from
the one prior to it, indicating a possibly continuously varying emission
source(s) with timescales ranging from months to one week or less.
Remarkably, the sense of the variability observed at shorter wavelengths
is reversed at longer wavelengths, with a constant pivot point
at $\lambda \sim 8.5$ \micron.
The largest flux difference was 60\% at short wavelengths and
30\% at long wavelengths, occurring over a period of 4 months between
epochs 2 and 3.  Another impressive change ($20-30$\%)
occurred over just one week between epochs 1 and 2.
The flux variations are fairly constant as a function of wavelength within
the intervals 5 to 7 and 11 to 35 \micron, respectively.  This suggests that
the continuum variability is tracing dust within relatively small temperature
ranges, and hence, localized regions within the disk.

\begin{figure}
\plotone{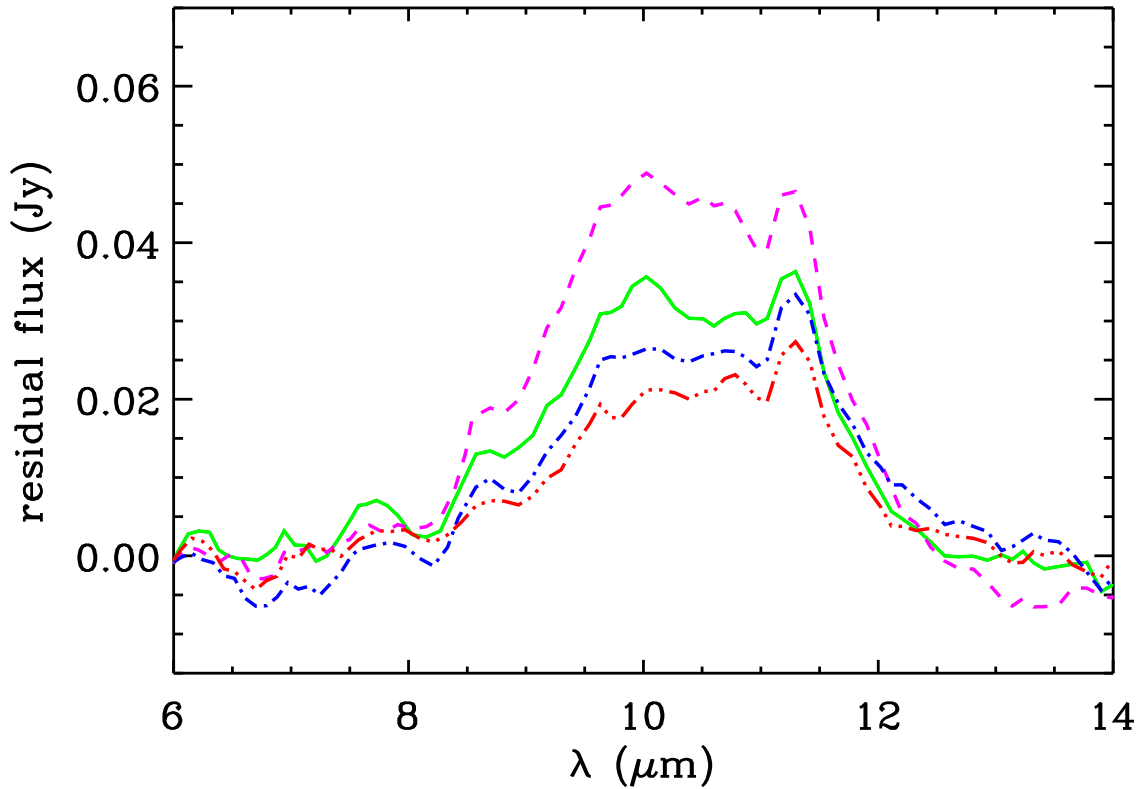}
\caption{Variation of the 10 \microns silicate emission.  The line scheme is
the same as in Figure~\ref{irs}.  The continuum has been subtracted
from each spectrum by fitting a polynomial to portions of the spectrum
containing no obvious features.  The emission features at 6.2, 7.7, 8.6,
and 11.2 \microns are PAH features produced either in the local ISM
(perhaps the result of background subtraction residuals) or the outer regions
of the LRLL 31 disk; they do not appear to vary significantly.
\label{silem}}
\end{figure}

In addition to the continuum emission, the 10 \microns silicate emission
feature also varied, in the same sense as the long-wavelength continuum
(Fig.~\ref{silem}).
The feature's profile possibly changes along with
the strength, with a hint of a stronger peak at 10 \microns (perhaps due
to an increase in emission from interstellar medium-like dust grains)
when the flux is brighter.  There is weak
evidence for variation in the 18 \microns silicate emission feature as well
(note the slight bumps in two of the difference spectra from about 15 to
25 \microns in Fig.~\ref{irs}); however, the signal-to-noise
is too low to yield a robust detection.

\begin{figure}
\plotone{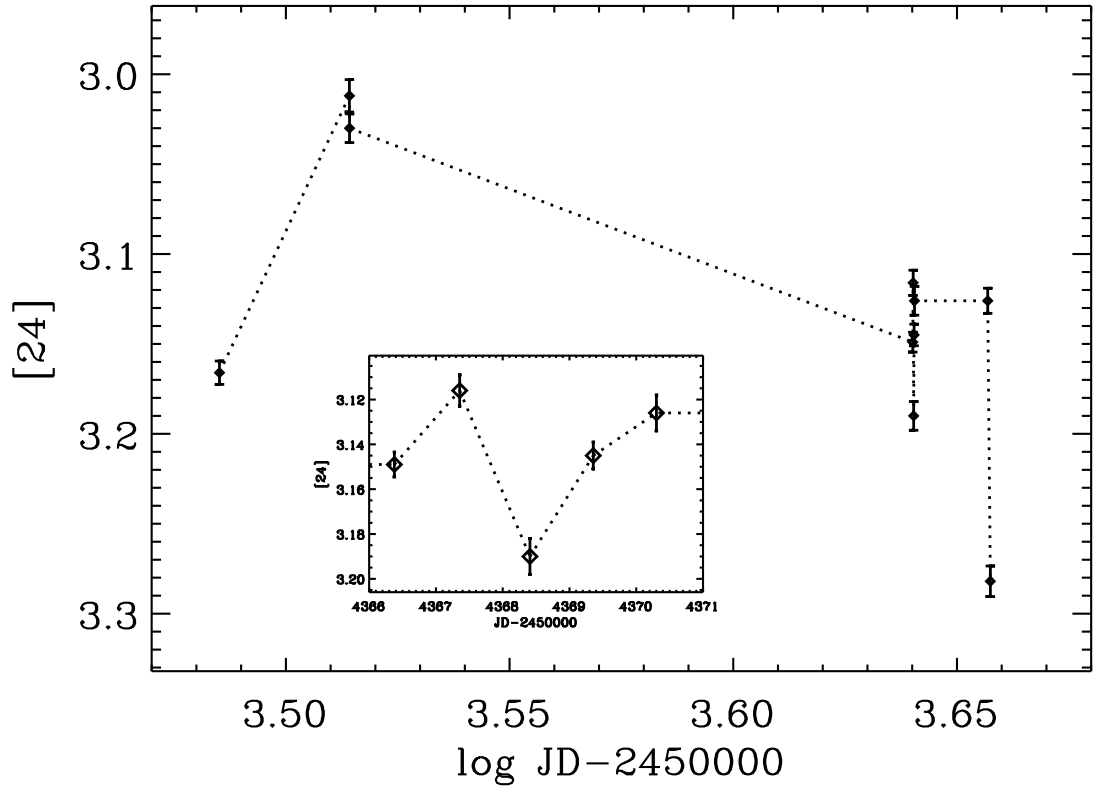}
\caption{Photometry time series at 24 \micron, in magnitudes.  The inset
shows a magnification of the 5 consecutive day data set.
\label{lcurve}}
\end{figure}

The MIPS images provide a finer time sampling of the flux at 24 \microns
(Fig.~\ref{lcurve}).  We find
flux variations over the full range of timescales, from as long as two years
to as short as one day.  The observations span a range in flux of about
0.3 magnitudes, or $\sim 25$\%, similar to the range exhibited by the
IRS spectra at $\lambda \gtrsim 10$ \micron, with an rms dispersion of 0.08
magnitudes (significantly larger than the typical measurement errors of
$\lesssim 0.02$ magnitudes).  A periodogram analysis detected no significant
periodicity, however, the sparse sampling prevents a quantitative assessment.

\begin{figure}
\plotone{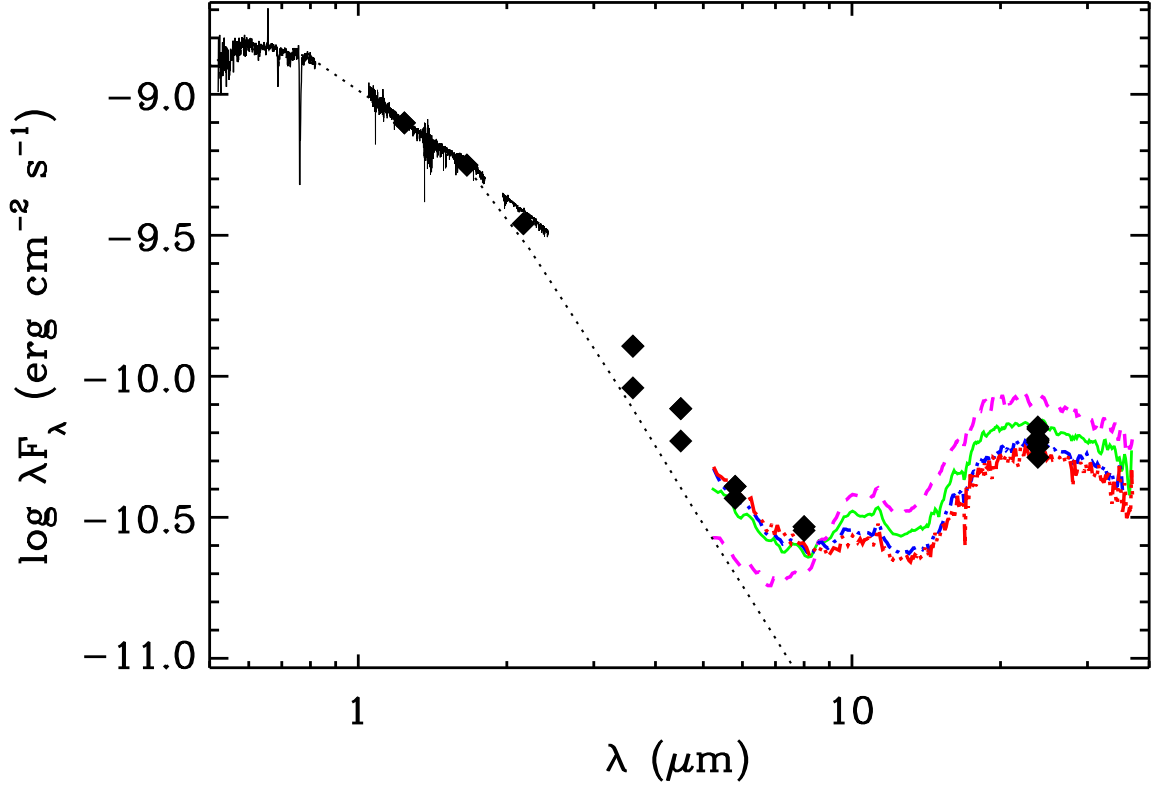}
\caption{Spectral energy distribution of LRLL 31.  The dereddened broadband
photometry (using $A_V = 10$ and reddening law from Mathis 1990 and
Flaherty et al. 2007; black diamonds) is from the photometric
observations listed in Table~\ref{obs}.  The stellar photosphere,
using empirical colors for a G6 star (Kenyon \& Hartmann 1995)
and scaled to the dereddened J-band flux, is indicated with the dotted line.
The solid line at 0.5-0.8 \microns shows the SPOL
spectrum, dereddened by $A_V=8.3$ and scaled to the stellar photosphere;
the visible absorption features are all telluric absorption bands.
The solid line at 1-2.4 \microns shows the SpeX SXD spectrum, dereddened
by $A_V=10$ and scaled to the J-band flux.
The Spitzer IRS spectra are shown with the same line scheme as in
Figure~\ref{irs}, all dereddened by $A_V = 10$.
\label{sed}}
\end{figure}

Figure~\ref{sed} shows the complete SED
for LRLL 31, with all epochs of data obtained to date.
The strength of the excess at longer wavelengths is typical of T Tauri-type
accretion disks, and indicates the presence of a gas-rich optically thick
disk at distances $\gtrsim 10$ AU from the star (a more precise location
would require detailed radiative transfer modeling of the SED).
Our detections of moderate
accretion lend further support, albeit indirectly, for such a gas reservoir.
The characteristic SED dip at 7-10 \microns and weak excess at shorter
wavelengths suggests a lack of warm dust in the inner parts of the disk,
indicating a large gap or inner hole in its dust distribution.
Such objects are often referred to as "transition" disks (Strom et al. 1989),
as they are believed to represent an evolutionary
stage where the disk is beginning to dissipate.  The (small) short-wavelength
infrared excess seen in LRLL 31 may be representative of either
optically thin dust entrained with gas accreting from the outer disk,
akin to known transition objects such as GM Aur and DM Tau
(Calvet et al. 2005), or an optically thick inner annulus akin to
the gapped ``pre-transitional" disk LkCa 15 (Espaillat et al. 2007, 2008).

\section{Discussion}

The infrared flux changes in LRLL 31 probe time-dependent
processes and may provide further insight into the structure of
transition disks.  We propose that the properties of this system
can be explained by vertical variations of
an optically thick disk annulus located close to the star.
A change in the inner disk emitting area will result
in a corresponding change in the short-wavelength infrared excess.
A large gap in the dust distribution separates this annulus from
an optically thick outer disk that is producing the long-wavelength
emission.  If the vertical extent of the inner disk is high enough,
it could then shadow parts of the outer disk (e.g., Dullemond et al. 2001;
O'Sullivan et al. 2005), resulting in less excess
emission at the long wavelengths when the excess at shorter wavelengths
is larger.  The magnitude of the effect would depend on the degree of
the change in height, its azimuthal extent, the vertical thickness and
distance of the outer disk, and the orientation of the system to the line
of sight.  In particular, a nearly edge-on inclination (as suggested
by our data) would be more likely to produce measurable flux variations
because of occultation effects.

There are several possible sources for a variable inner disk height.
One is an inner disk wall at the dust sublimation radius that is
``puffed" by direct irradiation from the central star.  The height of
such a wall can change if the disk mass accretion rate varies with time,
which would change both the irradiation flux (a combination of $L_*$
and the accretion luminosity, $L_{acc}$, liberated as disk gas falls
onto the stellar surface) and the in situ disk gas density.
This geometry was recently proposed to explain years-long infrared
variations seen in several Herbig Ae stars (Sitko et al. 2008).
Short-timescale variability in stellar accretion
rates has been observed fairly regularly in T Tauri stars,
and our own results suggest that LRLL 31 may exhibit $\mdot$ variations
of up to a factor of 5 on a $\lesssim 1$ year timescale.  More measurements
with shorter cadences are needed to fully test this possibility.
Nevertheless, there are a number of problems with the puffed inner rim
hypothesis.  Irradiation variations are unlikely to be important since
$L_* > L_{acc}$ by a factor of 5-10 even at the highest measured value of
$L_{acc}$.  The presence of a close binary companion on an eccentric orbit
may provide an alternate source of variable illumination of the inner disk.
However, the almost non-existent short-wavelength excess emission seen
in our epoch 2 IRS spectrum is very difficult to explain with a sublimation
wall, which should always produce some amount of infrared emission.
An extreme drop in the disk density at the inner wall might render
the disk optically thin to the stellar radiation, but the week-long timescale
of the epoch 2 flux decline would require a very short-lived event
of this type and the mass accretion rate should be several orders of
magnitude lower than what we observed (also, such a drastic drop
has not been observed in any other T Tauri star).
 
A second possibility is that the inner disk is being dynamically
perturbed by a stellar or planetary companion.  Induced vertical variations
such as warps or spiral density waves would result in a change in emitting area
in the inner disk and enable shadowing of the outer disk.
The epoch 2 IRS spectrum could be explained with the right combination of
vertical geometry and edge-on inclination, for example a geometrically flat
disk with a warp whose irradiated side is obscured by the star and/or front
side of the disk.  This possibility is particularly exciting since
the presence of a companion may already be revealed indirectly by the disk gap
inferred from the SED shape.
The timescales for dynamical effects will be commensurate with
the orbital frequency of the perturber.  The shortest infrared variations
we see would require a perturber near the inner edge of the dust disk at
the sublimation radius ($\sim 0.2$ AU), where the orbital timescales are
about 3-4 weeks.  This estimate suggests that a consistent model may be possible
in which a companion has partially cleared the inner region of the disk
and continues to perturb the system.

Our results for this one object indicate that exciting discoveries
of protoplanetary disk structure may be made through multi-epoch
infrared observations.  Further monitoring of a larger sample of
young stellar objects is needed to quantify how frequent mid-infrared
variability may be, and what (if any)
are the typical amplitudes and timescales.  In the case of LRLL 31 itself,
future multi-epoch, multi-wavelength observations are crucial for testing
the scenarios we outlined above.  Particularly important would be
2-5 \microns spectra to distinguish between optically thin or thick emission
from the inner disk, radial velocity monitoring to search for close companions,
long-term infrared monitoring to look for periodicity in the dust
excess flux variations, and long-baseline infrared interferometry
to try to spatially resolve the inner disk region.

\acknowledgements

This work is based in part on observations made with the {\it Spitzer}
Space Telescope, which is operated by the Jet Propulsion Laboratory,
California Institute of Technology under NASA contract 1407.
Support for this work was provided by NASA through Contract Number 960785
issued by JPL/Caltech.  We acknowledge K. Luhman for assistance with
spectral typing the SpeX spectrum.

\begin{deluxetable}{lll}
\tabletypesize{\small}
\tablewidth{0pt}
\tablecaption{LRLL 31 observations\label{obs}}
\tablehead{
\colhead{filter$/$instrument} &
\colhead{date} &
\colhead{ref}}
\startdata
SPOL (spectrum) & 2007 Dec 13 & this work\\
SPOL (image) & 2007 Dec 16 & this work\\
$J, H, K_s$ & 1998 Oct 5 & 2MASS\\
SpeX SXD & 2005 Dec 29 & this work\\
IRAC (GTO) & 2004 Feb 11 & Lada et al. 2006\\
IRAC (c2d) & 2004 Sep 8 & Jorgensen et al. 2006\\
MIPS (GTO) & 2004 Feb 21 & Lada et al. 2006\\
MIPS (c2d) & 2004 Sep 19 & Rebull et al. 2007\\
MIPS & 2007 Sep 23-27 & this work\\
& 2008 Mar 12, 19 & \\
IRS & 2007 Oct 9, 16 & this work\\
& 2008 Feb 24, Mar 2 & \\
\enddata
\end{deluxetable}

\end{document}